\documentclass[aps,twocolumn,showpacs,floats,superscriptaddress]{revtex4}
\usepackage{graphicx}
\usepackage{amssymb,amsmath}
\begin{document}

\title{Sign-Alternating Interaction Mediated by Strongly-Correlated Lattice Bosons}

\author{\c{S}.G. S\"{o}yler$^*$}
\affiliation{Department of Physics, University of Massachusetts,
Amherst, MA 01003, USA}

\author{B. Capogrosso-Sansone$^*$}
\affiliation{Department of Physics, University of Massachusetts,
Amherst, MA 01003, USA} \affiliation{Institute for Theoretical
Atomic, Molecular and Optical Physics, Harvard-Smithsonian Center
of Astrophysics, Cambridge, MA, 02138}

\author{N.V. Prokof'ev}
\affiliation{Department of Physics, University of Massachusetts,
Amherst, MA 01003, USA}  \affiliation{Russian Research Center ``Kurchatov
Institute'', 123182 Moscow, Russia}

\author{B.V. Svistunov}
\affiliation{Department of Physics, University of Massachusetts,
Amherst, MA 01003, USA} \affiliation{Russian Research Center
``Kurchatov Institute'', 123182 Moscow, Russia}
\begin{abstract}
We reveal a generic mechanism of generating sign-alternating inter-site interactions mediated
by strongly correlated lattice bosons. The ground state phase diagram of the
two-component hard-core Bose-Hubbard model on a square lattice at half-integer filling
factor for each component, obtained by worm algorithm Monte Carlo simulations,
is strongly modified by these interactions and features the solid+superfluid phase
for strong anisotropy between the hopping amplitudes. The new phase is a direct consequence
of the effective nearest-neighbor repulsion between ``heavy'' atoms mediated by
the ``light'' superfluid component. Due to their sign-alternating character,
mediated interactions lead to a rich variety of yet to be discovered quantum phases.
\end{abstract}
\pacs{03.75.Mn, 03.75.Hh, 67.85.-d, 05.30.Jp} \maketitle


The first proposal for studying models of strongly correlated systems with cold
atoms in optical lattices was put forward a decade ago~\cite{Jaksch1}. Since then,
control over lattice geometry and interaction strength has
increased dramatically, opening up new directions in the study of
quantum phases of cold gases. (For reviews,
see~\cite{Bloch,Lewenstein}.) Thanks to refinements in
experimental and theoretical tools, it is now possible to look at
exotic quantum states which arise in bosonic systems with pseudospin degrees
of freedom or multiple species.
In the realm of two-component systems, one goal is to
realize models of quantum magnetism by using hyperfine
states of an atom~\cite{hyperfine}. By controlling
superexchange interactions of particles confined in an optical
lattice, it is possible to switch between different ground
states~\cite{seperexchange interactions}.
Another important development is experimental realization of heteronuclear
bosonic mixtures of ${}^{87}\textrm{Rb}-{}^{41}\textrm{K}$
in a three-dimensional optical lattice~\cite{mixtures in OL}.
Moreover, Ref.~\cite{controlled interactions} reports results
for fine control over interspecies scattering length, including the zero-crossing point.
These achievements indicate that two-component systems in optical lattices
with tunable interspecies interaction via Feshbach resonances are within the
reach of current experiments. The two-component
2D bosonic system is also in the focus of Optical Lattice Emulator project
supported by DARPA and aimed at the development, within the next few years,
of experimental tools of accurately mapping phase diagrams of lattice systems by
emulating them with ultracold atoms in optical lattices.

Experimental studies of lattice solids---states with broken translation symmetry---are intriguing and fundamentally important, especially in 2D. For the prominent example, we refer to the problem
of deconfined criticality proposed for the solid-to-superfluid quantum phase transition in 2D.
The matter of interdisciplinary interest here is to validate the idea of a (hidden) duality
between the superfluid and solid orders leading to a conceptually new criticality \cite{DCP}.
Lattice solids also offer the possibility of having a supersolid phase
featuring both broken translation symmetry and the ability to support a superflow, e.g. in
a single-species square-lattice bosonic system with soft-core on-site interactions
{\it and} appropriately strong nearest-neighbor interactions \cite{supersolid}.

Solid phases in the single-species bosonic system require going beyond the on-site interaction.
The standard Bose-Hubbard model \cite{Jaksch1} supports only two phases: a superfluid (SF) and a Mott insulator (MI); the latter is not a solid because it lacks the broken translation symmetry.
A considerable theoretical effort has been made to understand how inter-site interactions can be generated in atomic gases. One proposal is to use cold polar molecules \cite{polar1} featuring long range dipole-dipole  interactions (see also review  \cite{polar1}, and references therein).  In Refs.~\cite{Buchler}, the authors suggest a technique for tuning the shape of long-range
interactions between polar molecules by applying static and microwave fields. Another, experimentally more challenging, proposal is to excite atoms to higher bands in an optical lattice \cite{excited}.

Within this framework, the two-component bosonic system with purely on-site interactions
is a reasonable alternative route to obtain exotic single-species systems
(at present, it is hard to reach low temperatures with lattice fermions).
At a commensurate filling and strong enough interaction, opening a gap in the net-charge sector, the two-component mixture becomes equivalent to a single-component system with nearest-neighbor interactions, describing the iso-spin sector \cite{Boninsegni,Svistunov-Kuklov,Duan-Demler-Lukin,Demler-Lukin}.
The checker-board (CB) solid phase arising in this case is equivalent to the
N\'{e}el antiferromagnet \cite{Boninsegni,Svistunov-Kuklov,Demler-Lukin}. The CB-SF quantum phase transition (with respect to the original components, the SF state is a super-counter-fluid (SCF)  \cite{Svistunov-Kuklov}, or, equivalently, planar ferromagnet in iso-spin terminology \cite{Svistunov-Kuklov,Demler-Lukin}) is known to be of the first order. 
Unfortunately, the supersolid phase is not predicted in this parameter regime.  
There have been extensive theoretical studies of quantum phases in
two-component systems  \cite{Svistunov-Kuklov,Svistunov-Kuklov-Prokofev,Demler-Lukin,Duan-Demler-Lukin,Isacsson}, but, to the best of our knowledge, all studies overlooked the possibility of having various solids in the {\it absence}
of the net-charge localization. Namely, they missed a generic mechanism of
inducing inter-site sign-alternating interactions by strongly-correlated {\it bosonic} environment (cf.~\cite{Viverit}), analogous to the RKKY (Ruderman-Kittel-Kasuya-Yosida) interaction mediated
by fermions \cite{RKKY}.

In this Letter, we reveal and quantify the mechanism of the mediated sign-alternating interactions,
and discuss it in the context of the ground state phase diagram for
\textit{hard-core} bosons with repulsive interspecies interaction at half filling for each component:
\begin{equation}
H=-\sum_{<ij>} \left( t_{\rm a} a^{\dag}_i\,a_j + t_{\rm b} b^{\dag}_i\,b_j  \right)
+U\sum_{i}n^{\rm (a)}_in^{\rm (b)}_i \; ,
\label{hamiltonian}
\end{equation}
where, $a^{\dag}_i(a_i)$ and $b^{\dag}_i(b_i)$ are bosonic
creation (annihilation) operators, $t_{\rm{a}}$ and $t_{\rm{b}}$ are hopping
matrix elements between the nearest neighbor sites for two species of bosons ($A$ and $B$) on a simple
square lattice with $N=L \times L$ sites,
and $n_i^{\rm (a)} = a^{\dag}_i a_i$, $n_i^{\rm (b)}= b_i^{\dag}b_i$.
This model can be implemented experimentally \cite{controlled interactions} and is considered to be
the simplest one with purely contact interactions and yet highly nontrivial phase diagram.

Model (\ref{hamiltonian}) was studied previously using
a combination of variational and mean field theories \cite{Demler-Lukin}
which, in general, can not guarantee the accuracy of results. With
Monte Carlo (MC) simulations by Worm Algorithm  \cite{Worm}, we obtain
the first precise data for the ground state phase diagram.
For weak anisotropy between $t_{\rm{a}}$ and $t_{\rm{b}}$ and large $U$,
our results confirm the basic phases and
transitions between them proposed in Ref.~\cite{Demler-Lukin}. We, however, find strong quantitative
differences (up to $50\%$ to $100\%$) in the location of transition lines. For large anisotropy and moderate-to-weak
interactions we find a completely new structure of the phase diagram. It is shaped by the effective
Hamiltonian obtained for the ``heavy'' (small hopping) component
after the ``light'' component is integrated out.
The resulting nearest-neighbor and longer-range interactions (similar to the effective potential
between the ions in solids mediated by electrons)
stabilize the checker-board (CB) solid phase of heavy atoms for sufficiently strong anisotropy between
$t_{\rm{b}}$ and $t_{\rm{a}}$.
A  surprising result of the present study is that effective mediated interactions are
oscillating from strong on-site attraction to much weaker nearest neighbor repulsion and back to
a tiny attractive tail.
In a broad perspective, this type of mediated interactions will result in
interesting solid, and guaranteed supersolid \cite{supersolid} orders in related models.
Moreover, for soft-core bosons, one can look for phases and phase transitions
which involve multi-particle bound states and order parameters (``multi-mers'').

Before we discuss our findings in more detail let us review the key phases and limiting cases of
model (\ref{hamiltonian}). In the strong coupling limit,  $U \gg t_{\rm{a}},t_{\rm{b}}$,
 it can be mapped (within the second-order perturbation theory) onto the spin-1/2 Hamiltonian (see
e.g.~\cite{Boninsegni,Svistunov-Kuklov,Demler-Lukin})
$H_{XXZ} = \sum_{<ij>} [ -J_{xy} (\sigma_j^x \sigma_i^x + \sigma_j^y \sigma_i^y)+J_z \sigma_j^z \sigma_i^z]$
with positive $J_{xy}, J_z \sim t^2/U$.
The latter features two possible ground states: (i) an antiferromagnetic state
with \textit{z}-N\'{e}el order for $J_z>J_{xy}$, and (i) an \textit{XY}-ferromagnetic
state for $J_{xy}>J_z$. In bosonic language,
the \textit{z}-N\'{e}el state corresponds to the CB solid order for both A- and B-particles
(we will abbreviate it as 2CB).
It is characterized by non-zero structure factor
$S_{\rm{a,b}}(\textbf{k})=N^{-1} \sum_{\textbf{r}} \exp\left[i
\textbf{k}\textbf{r}\right] \langle n^{\rm{(a,b)}}_0 n_{\textbf{r}}^{\rm{(a,b)}}\rangle$.
The \textit{XY}-state is representing the SCF phase featuring an order parameter
$\left\langle a^{\dag}b\right\rangle$. Both 2CB and SCF have to be regarded as Mott
insulators as far as the total number of particles is concerned, i.e. there exist a finite gap to
dope the system. Thus only counter-propagating A- and B-currents with the zero net particle flux posses
superfluid properties in the SCF state.
Under the mapping one finds that $J_z>J_{xy}$ everywhere except at $t_{\rm a}=t_{\rm b}$ when the spin Hamiltonian
becomes SU(2)-symmetric. Thus higher-order symmetry-breaking terms are necessary to decide which phase,
2CB or SCF, survives. Ref.~\cite{Demler-Lukin} provided a variational argument showing that SCF is stabilized in the vicinity of the $t_{\rm a}=t_{\rm b}$ line, and our data unambiguously confirm the
validity of this conclusion.

At weak inter-species interaction, $U \ll t_{\rm{a}},t_{\rm{b}}$, we expect that the
ground state is that of two miscible strongly interacting (due to hard-core intra-component repulsion)
superfluids (2SF). Finally, for $t_{\rm{b}} \ll U \le t_{\rm{a}}$ we should have a phase where
B-particles form the CB solid if effective interactions mediated by the superfluid A-component
are repulsive and short-ranged (we abbreviate it as CB+SF).

\begin{figure}[t!]
\hspace*{-0.5cm} \centerline{\includegraphics[angle=0,width=3.6in
]{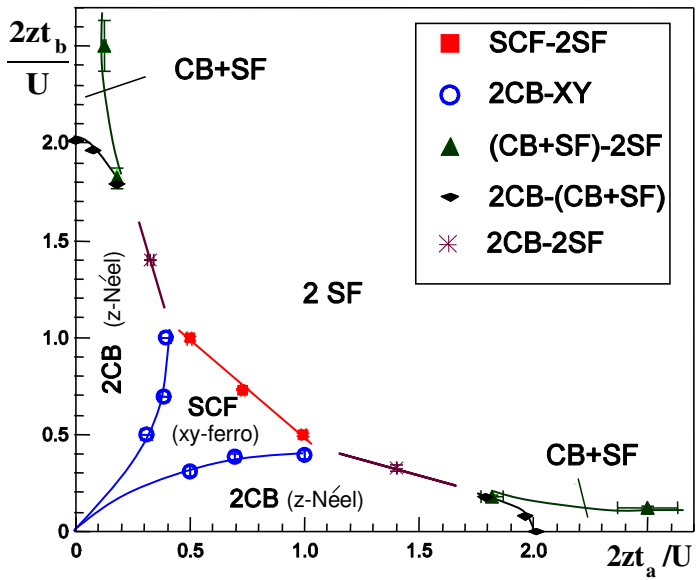}} \caption{(Color online). Phase diagram of
model~(\ref{hamiltonian}) on a square lattice at half-integer
filling factor for each component. The observed transition lines are
2CB-SCF (first-order),
SCF-2SF (second-order),
2CB-2SF (first-order),
2CB-CB+SF (second-order), and
CB+SF-2SF (first-order). Lines are used to guide an eye.}
\label{fig1}
\end{figure}

Our simulation method is based on the lattice path integral representation
and Worm Algorithm \cite{Worm}. The original version was generalized to deal with
two-component systems following ideas introduced for classical $j$-current models
\cite{SCF-PSF}.  The simulation configuration space now includes the possibility
of having two types of disconnected worldlines (worms) representing off-diagonal correlation
functions (Green's functions). In order to allow efficient sampling of the SCF phase
(any paired phases for that matter)
it is necessary to enlarge the configuration space and consider worldline trajectories
with two worms propagating simultaneously.
The results for the phase diagram are summarized in
Fig.~\ref{fig1}. To detect the SCF phase we have
calculated the stiffness of the \textit{relative} superfluid flow from the standard
winding number formula \cite{Pollock} $\rho_{\rm{SCF}} = \beta^{-1} \left\langle (W_{\rm{a}} - W_{\rm{b}})^2
\right\rangle $, where $W_{\rm{a(b)}}$ are winding
numbers of worldlines $A(B)$, and $ \beta $ is the inverse
temperature. In SCF the sum of winding numbers is zero in the thermodynamic limit.
We confirm the SCF ground state for $t_{\rm{a}}\sim t_{\rm{b}}$ and sufficiently strong interactions.
It survives at arbitrary large $U$ along the diagonal $t_{\rm{a}} = t_{\rm{b}}$,
directly demonstrating that higher order terms in the effective
spin-1/2 Hamiltonian break the SU(2) symmetry in favor of the {\textit XY}-order.
To locate the weakly first-order superfluid-solid 2CB-SCF line (circles in Fig.~\ref{fig1})
we have used the flowgram method which works well for both first and second order transitions,
and is particularly helpful for telling the former from the latter (see Ref.~\cite{flowgram} for details).

Though the 2CB-SCF transition is expected to be first-order, one may not
exclude the possibility of the intermediate supersolid phase. We did search for evidence of
the state featuring both the SCF and CB orders but did not find any.
On the contrary, we observed phase coexistence which brings us to the conclusion
that the 2CB-SCF line remains un-split.

\begin{figure}[t!]
\vspace*{-0.30cm} \hspace*{-0cm}
\centerline{\includegraphics[angle=-90, width=3.5in, keepaspectratio=true]{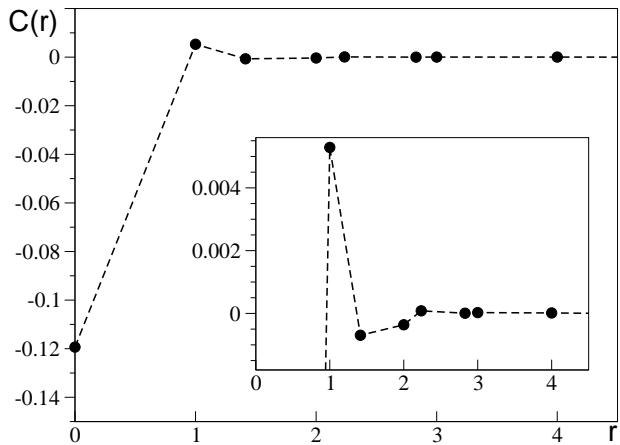}}
\caption{The $C(r)$ function in a system of light hard-core bosons at half-integer filling; the distance $r$ is measured in units of lattice spacing.
The calculation was done for the $L\times L = 10\times 10$ system
at low temperature. The nearest neighbor repulsion is clearly visible,
though the overall strength of the effective coupling is very small. Error bars are smaller than
symbol size. }\label{fig2}
\end{figure}

The continuous SCF-2SF transition (squares in Fig.~\ref{fig1}), is expected to be
in the $(d+1)$-dimensional U(1) universality class
characteristic of the MI-SF quantum phase transition \cite{SCF-PSF} .
As the system crosses into the 2SF phase, it develops single-component
order parameters $\left\langle a \right\rangle \neq 0$ and $\left\langle b \right\rangle \neq 0$
along with the non-zero superfluid stiffness in the total winding number channel,
$\rho_{2SF} = \beta^{-1} \langle ( W_{\rm{a}} +
W_{\rm{b}}) ^2 \rangle$. To locate the transition line precisely
we have employed standard finite size scaling
arguments and extracted the critical point from the intersection
of $\rho_{2SF}(U/t_{\rm{a,b}})L$ curves calculated for different
system sizes $L$ at $\beta \propto L$.

The SCF phase disappears for $U/t_{\rm{a,b}}\lesssim
8$. In this parameter region, the system undergoes the first-order
2CB-2SF transition (stars in Fig.~\ref{fig1}) up
to $U/t_{\rm{a,b}}\simeq 4$ where the 2CB phase disappears.

Not too surprisingly for the 2D case, the mean-field and variational treatments turn
out to rather inaccurate quantitatively: The actual transition lines
are 50\% to 100\% away from the predicted ones. For
comparison, in Fig.~\ref{fig1} we have used the same units as in
Ref.~\cite{Demler-Lukin} ($z=4$ is the coordination number on the square lattice).
Even for strong inter-species interaction $U/t_{\rm{a,b}}\sim 16$ we do
not find good quantitative agreement.

We now turn to the most interesting, for purposes of this Letter, region of the phase diagram
with strong anisotropy between hopping amplitudes. In this regime one of the
components is much heavier (let it be component B) than the other.
As the hopping amplitude $t_{\rm{a}}$ increases, the light component undergoes
a second-order MI-SF transition in $(d+1)$-dimensional U(1) universality class
(diamonds in Fig.~\ref{fig1}). [Since translation invariance is broken by the CB order
of B-particles, the filling factor of A-particles is unity per unit cell.]
Beyond this transition, the superfluid A-bosons provide an effective interaction for B-bosons
which for sufficiently small $t_{\rm{b}}$ stabilizes the CB order in the heavy component.
Though formally CB+SF breaks both translation and gauge symmetries it can not be called
a supersolid because the density wave in the A-component is induced from ``outside'' by an insulating
heavy phase (this is reminiscent of conventional solid-state superconductors). However, in cold
atomic systems with $A$ and $B$ particles referring to different hyperfine states of the same atom
we  have an experimental ``knob'' to render the CB+SF phase a genuine supersolid by inducing
an arbitrary small hybridizing interaction $g (a^{\dag} b +H.c.)$.

To describe the CB+SF phase semi-quantitatively let us look at the limit
$t_{\rm{b}}\ll U \ll t_{\rm{a}}$. Within the lowest-order perturbation theory
in $U$ an effective mediated interaction between heavy atoms is pairwise and
of the order of $U^2/zt_{\rm{a}}$. Since the superfluid liquid of hard-core A-particles is
strongly correlated, the actual strength and form of $V^{\rm eff}_{ij}$
has to be computed numerically. We determine it from the exact (at $U/t_{\rm a} \to 0 $) relation
$V^{\rm eff} (r) = (U^2/t_{\rm a})\, C(r)$, where
\[
C(r)=\mathop {\lim }\limits_{U/t_{\rm a} \to 0 } \frac{n(r)-1/2}{U/t_{\rm a}} \
\]
is the scaled density profile in response to the heavy atom at the origin, and
$r$ is the dimensionless distance between the B-atoms.
Since $C(r)$ is expected to decay exponentially fast with $r$ \cite{remark},
the simulated system can be relatively small. The result is shown in Fig.~\ref{fig2}.
It turns out that $V^{\rm eff} (r)$ is more than an order of magnitude smaller
then a naive estimate $U^2/zt_a$. Moreover, it is sign-alternating, with strong on-site
attraction, much weaker nearest-neighbor repulsion, and nearly negligible attractive tail
for longer range interactions. For the hard-core model studied here, the dominant interaction is
the nearest-neighbor one, $V^{\rm eff}(1) \approx 5.4\times 10^{-3}\ U^2/t_{\rm a}$, explaining
the origin of the CB phase of heavy atoms.

On the basis of $V^{\rm eff}(1)$, the melting first-order (CB+SF)-2SF transition is predicted to occur
along the $t_{\rm{b}}\approx V^{\rm eff}(1)/2$, or $4z^2 t_{\rm{b}}t_{\rm{a}}/U^2 \approx 0.17$, line,
i.e. the CB+SF phase survives close to the vertical and horizontal axis
in Fig.~\ref{fig1} all the way to $t_{\rm{a},\rm{b}}/U \to \infty$.
The above asymptotic estimate is less than a factor of two smaller than the
data points in Fig.~\ref{fig1}, even though $U$ is still relatively large.

In conclusion, we have presented accurate results, based on path
integral Monte Carlo, for the phase diagram of the two component
hard core Bose-Hubbard model on a square lattice and half-integer
filling factor for each component. The system can be realized
experimentally with heteronuclear bosonic mixtures in optical
lattices with tunable interspecies interactions. We  reveal the existence of an additional
CB+SF state which radically changed the topology of the phase diagram.
The CB+SF phase, which exists for strong anisotropy between the
hopping amplitudes and weak enough interaction, is a direct consequence of
effective interactions mediated by the light, strongly correlated superfluid
component. Mediated interactions are sign-alternating,
and lead to exciting possibilities of realizing new quantum phases in the
two-species bosonic systems.

Many questions remain open. Finite
temperature properties and melting of the
\textit{z}-N\'{e}el/\textit{xy}-ferromagnet phases are of interest
in the study of quantum magnetism. Studies of soft core bosons are of special interest
because they admit the possibility of forming ``multi-mers'' due to strong mediated on-site
attraction. One may also study supersolid phases on square and triangular lattices, not to
mention the need for dealing with more realistic systems, i.e.
including effects of parabolic confinement and finite number of particles
as in experimental setups.

We thank L. Pollet, E. Demler, and M. Lukin for fruitful
discussions. This work was supported by ITAMP, DARPA OLE program
and NSF grant PHY-0653183.
\\$*$ These authors contributed equally.


\begin{thebibliography}{23}
\bibitem{Jaksch1} D. Jacksch, C. Bruder, J. I. Cirac, C. W. Gardiner, and P. Zoller, Phys. Rev. Lett. \textbf{81}, 3108 (1998).
\bibitem{Bloch} I. Bloch, J. Dalibard, W. Zwerger, Rev. Mod. Phys. \textbf{80}, 885, (2008).
\bibitem{Lewenstein} M. Lewenstein, A. Sanpera, V. Ahufinger, B. Damski, A. S. De, and U. Sen, Adv. Phys. \textbf{56}, 243 (2008).
\bibitem{hyperfine} S. F\"{o}lling, S. Trotzky, P. Cheinet, M. Feld, R. Saers, A. Widera, T. M\"{u}ller, and I. Bloch, Nature \textbf{448}, 1029 (2007); M. Anderlini, P. J. Lee, B. L. Brown, J. Sebby-Strabley, W. D. Phillips, and J. V. Porto, Nature \textbf{448}, 452 (2007).
\bibitem{seperexchange interactions} S. Trotzky, P. Cheinet, S. F\"{o}lling, M. Feld, U. Schnorrberger, A. M. Rey, A. Polkovnikov, E. A. Demler, M. D. Lukin, I. Bloch, Science \textbf{319}, 294 (2008).
\bibitem{mixtures in OL} J. Catani, L. De Sarlo, G. Barontini, F. Minardi, and M. Inguscio, Phys. Rev. A \textbf{77}, 011603(R) (2008).
\bibitem{controlled interactions} G. Thalhammer, G. Barontini, L. De Sarlo, J. Catani, F. Minardi, and M. Inguscio, cond-mat: 0803.2763.
\bibitem{DCP} T. Senthil, A. Vishwanath, L. Balents, S. Sachdev, and M. P. A. Fisher, Science \textbf{303}, 1490 (2004); T. Senthil, L. Balents, S. Sachdev, A. Vishwanath and M. P.A. Fisher, Phys. Rev. B \textbf{70}, 144407 (2004); O. I. Motrunich and A. Vishwanath, Phys. Rev. B \textbf{70}, 075104 (2004).
\bibitem{Boninsegni} M. Boninsegni, Phys. Rev. Lett. \textbf{87}, 087201 (2001).
\bibitem{Svistunov-Kuklov} A. B. Kuklov and B. V. Svistunov, Phys. Rev. Lett. \textbf{90}, 100401 (2003).
\bibitem{Duan-Demler-Lukin} L. M. Duan, E. Demler, and M. Lukin, Phys. Rev. Lett. \textbf{91}, 090402 (2003).
\bibitem{Demler-Lukin} E. Altman, W. Hofstetter, E. Demler, and  M. Lukin, New J. Phys. \textbf{5}, 113, (2003).
\bibitem{supersolid} P. Sengupta, L. P. Pryadko, F. Alet, M. Troyer and G. Schmid, Phys. Rev. Lett. \textbf{94}, 207202 (2005).
\bibitem{polar1} K. Goral, L. Santos, M. Lewenstein, Phys. Rev. Lett. \textbf{88}, 170406 (2002).
\bibitem{polar2} G. Pupillo, A. Micheli, H. P. B\"uchler, P. Zoller, in: {\it Cold Molecules: Creation and Applications},  eds.  R.V. Krems, B. Friedrich, and W.C. Stwalley, (Taylor and Francis, 2008).
\bibitem{Buchler} H. P. B\"{u}chler, E. Demler, M. Lukin, A. Micheli, N. Prokof'ev, G. Pupillo and P. Zoller, Phys. Rev. Lett. \textbf{98}, 060404 (2007);  G. Pupillo, A. Micheli, H. P. B\"uchler and P. Zoller, Phys. Rev. A \textbf{76}, 043604 (2007).
\bibitem{excited} V. W. Scarola and S. Das Sarma, Phys. Rev. Lett. \textbf{95}, 033003 (2005)
\bibitem{Svistunov-Kuklov-Prokofev} A. Kuklov, N. Prokof'ev, and B. Svistunov, Phys. Rev. Lett. \textbf{92}, 050402 (2004).
\bibitem{Isacsson} A. Isacsson, M. C. Cha, K. Sengupta, and S. M. Girvin, Phys. Rev. B \textbf{72} 184507 (2005).
\bibitem{Viverit}  H. Heiselberg, C. J. Pethick, H. Smith, and L. Viverit,
 Phys. Rev. Lett. \textbf{85}, 2418 (2000).
\bibitem{RKKY} J.H. Van Vleck, Rev. Mod. Phys. {\bf 34}, 681 (1962), and references therein.
\bibitem{Worm} N. V. Prokof'ev, B.V. Svistunov, and I.S. Tupitsyn, Phys. Lett. A \textbf{238}, 253 (1998); Sov. Phys. JETP \textbf{87}, 310 (1998).
\bibitem{SCF-PSF} A. Kuklov, N. Prokof'ev, and B. Svistunov, Phys. Rev. Lett. \textbf{92}, 030403 (2004).
\bibitem{Pollock} E. L. Pollock and D. M. Ceperley, Phys. Rev. B \textbf{36}, 8343 (1987).
\bibitem{flowgram} A. B. Kuklov, N. V. Prokof'ev, B. V. Svistunov, and M. Troyer, Ann. of Phys. \textbf{321}, 1602 (2006).
\bibitem{remark} Static perturbation of density,
is not a Goldstone mode, and thus cannot demonstrate a power-low decay.
\end{thebibliography}
\end{document}